\documentclass[conference]{IEEEtran}
\newif\ifsubmission
\submissionfalse

\usepackage{adjustbox}
\usepackage{amsmath,amssymb,amsfonts}
\usepackage{algorithmic}
\usepackage{balance}
\usepackage{booktabs}
\usepackage[font={footnotesize}]{caption}
\usepackage{cite}
\usepackage{color, colortbl}
\usepackage{enumitem}
\usepackage{graphicx}
\usepackage[breaklinks,hidelinks]{hyperref}
\usepackage{cleveref} %
\usepackage{listings}
\usepackage{mdframed}
\usepackage{multirow}
\usepackage[numbers,sort]{natbib}
  \def\BibTeX{{\rm B\kern-.05em{\sc i\kern-.025em b}\kern-.08em
    T\kern-.1667em\lower.7ex\hbox{E}\kern-.125emX}}

\usepackage{pdfcomment}
\usepackage{relsize}
\usepackage[nolist]{acronym}
\usepackage{siunitx}
\usepackage{soul}
\usepackage{subcaption}
\usepackage{tabularx}
\usepackage{textcomp}
\usepackage{wasysym}
\usepackage{xcolor}
\usepackage{xspace}

\newcommand{\head}[1]{\textnormal{\textbf{#1}}}

\def\Snospace~{\S{}}

\renewcommand{\paragraph}[1]{{\vskip 6pt \noindent\textbf{#1.} }}

\newcommand{\ie}{i.e.,\xspace} %

\newcommand{\perc}[1]{%
  \ifstrempty{#1}%
  {\,\si{\percent}}%
  {\SI{#1}{\percent}}%
}

\newcommand{\ember}{Ember\xspace}

\newcommand{\lgbm}{LightGBM\xspace}

\newcolumntype{L}[1]{>{\raggedright\arraybackslash}p{#1}}

\newcommand{\defname}{PEberus\xspace}
\newcommand{\defnameit}{\textit{PEberus}\xspace}

\begin{document}

\begin{acronym}
\acro{pe}[PE]{Portable Executable}
\acro{gbdt}[GBDT]{Gradient Boosted Decision Tree}
\acro{coff}[COFF]{Common Object File Format}
\end{acronym}

\title{Against All Odds: Winning the Defense Challenge in an
Evasion Competition with Diversification}

 \author{
	\IEEEauthorblockN{
		Erwin Quiring,
		Lukas Pirch,
		Michael Reimsbach,
		Daniel Arp,
		Konrad Rieck
	}
	\vspace{0.2cm}
	\IEEEauthorblockA{Technische Universit\"at 
	Braunschweig\\Braunschweig, Germany}
}

\maketitle
\ifsubmission\else
  \thispagestyle{plain}
  \pagestyle{plain}
\fi

\begin{abstract}
Machine learning-based systems for malware detection operate in a 
hostile environment. 
Consequently, adversaries will also target the learning system and use 
evasion attacks to bypass the detection of malware. 
In this paper, we outline our learning-based system \defnameit that
got the first place in the defender challenge of the Microsoft Evasion
Competition, resisting a variety of attacks from independent
attackers.  Our system combines multiple, diverse defenses: we address
the semantic gap, use various classification models, and apply a
stateful defense.  This competition gives us the unique opportunity to
examine evasion attacks under a realistic scenario. It also highlights
that existing machine learning methods can be hardened against attacks
by thoroughly analyzing the attack surface and implementing
concepts from adversarial learning.  Our defense can serve as an
additional baseline in the future to strengthen the research on secure
learning\footnote{The defense is publicly available at
  \url{https://github.com/EQuiw/2020-evasion-competition} }.

\end{abstract}

\section{Introduction}

Machine learning is a powerful tool for detecting malware. The 
capability to automatically infer and generalize patterns from data 
allows detecting newly emerging malware.
However, machine learning itself introduces a considerable attack 
surface, as previous work in adversarial learning unveils.  
Possible attacks range from exploiting the preprocessing
stage~\citep{QuiKleArpJohRie18, XiaCheShe+19}, poisoning the training
data~\citep[e.g.,][]{BigNelLas11, GuDolGar17, LiuMaAaf+18}, stealing
the model~\citep{TraZhaJuel+16}, to misleading the
prediction~\citep[e.g.,][]{BigCorMai+13, CarWag17}.

As a result, it is vitally important to consider machine
learning-related attacks in addition to the underlying security
problem. The 2020 Machine Learning Security Evasion 
Competition~\citep{web:Contest} focuses
on the threat of evasion attacks with Windows \ac{pe} malware. This
competition provides a unique opportunity: It allows researchers to 
take the role of a defender or attacker in a real scenario without 
perfect knowledge. Defenses can be examined against evasion attacks 
from independent real-world attackers.

Our defense \defnameit got the first place in the defender challenge, 
resisting a variety of sophisticated attacks. Our solution is based on 
diversification and consists of three main concepts.
First, various heuristics address the semantic gap. This gap between 
the semantics of a \ac{pe} program and its feature representation 
allows relatively simple functionality-preserving attacks.
Second, multiple classification models use distinct feature sets to
classify malware reliably while considering targeted attacks against
the model. A stateful defense finally detects iterative attacks that
exploit the API access to the classifier.
Although our solution fends off the majority of attacks in the
competition, it is limited to static analysis, and thus a few attacks
based on obfuscation succeeded. Note that the use-case is Windows
\ac{pe} malware, but our insights and concepts are generally usable
for other security domains~as~well.

In this paper, we outline the design of our defense. After
providing background information about the contest and \ac{pe} malware
detection (\autoref{sec:background}), we analyze the attack surface of
machine learning-based malware detection from a practical and
theoretical point of view (\autoref{sec:attacks}). This provides the
basis for our defense in \autoref{sec:defenses}. We describe the
results and insights from the competition in \autoref{sec:eval}, and 
conclude the paper in \autoref{sec:conclusion}.

\section{Background}\label{sec:background}
Before examining the attack surface and our defenses, we briefly 
introduce background information on the competition, the structure 
of Windows PE files, existing approaches to detect malware, and the 
threat of evasion attacks.

\subsection{Microsoft's Evasion Competition}
\label{subsec:competition}
The Machine Learning Security Evasion Competition by 
Microsoft~\citep{web:Contest} focuses on the robustness of machine 
learning against evasion attacks in the context of malware detection.
The contest is structured in two challenges.

We participated in the \emph{defender challenge}. The participants
develop their own solution to detect Windows \ac{pe} malware without
any restrictions on the used dataset, features, or learning method.
Yet, a valid solution needs to fulfill three requirements: (a) the
maximum false positive rate is \perc{1}, (b) the minimum true positive
rate is \perc{90}, (c) the system must report a decision within five
seconds. The competition organizers regularly check these conditions
on a holdout set that is unknown to the participants.

In the subsequent \emph{attacker challenge}, participants need 
to bypass the detection of the developed defenses from the first 
round for 50~Windows malware samples. Only a black-box API access 
with a binary output is provided without further knowledge about the 
defenses. The attackers have any freedom to manipulate the \ac{pe} 
file, but with the requirement to preserve the functionality.
This is tested by a dynamic analysis for the submitted adversarial 
samples. The maximum \ac{pe} file size for all submissions is 2~MiB.

\subsection{Windows PE File Structure}
\label{subsec:pe_file}

\begin{figure}
	\centering
	\includegraphics{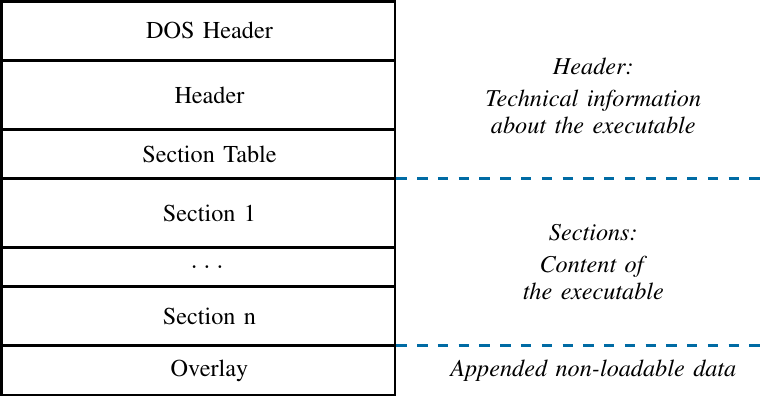}
	\caption{Overview about the \ac{pe} File Structure}
	\label{fig:pe_file_structure}
\end{figure}

The detection of \ac{pe} malware requires a basic understanding of the 
file format. 
\autoref{fig:pe_file_structure} illustrates the general \ac{pe} 
structure. It can be divided into three parts: the header, the 
sections, and the overlay. All \ac{pe} files start with a DOS header. 
Another header follows that contains the 
actual information for a PE file, such as the type of the target 
machine (e.g., x64) or a timestamp of creation.
The section table describes each section, including its size and 
address in the \ac{pe} file and memory. Padding can be necessary to 
meet the file alignment, resulting in a slack space between 
sections.
A section can be, for instance, executable code, an export or import 
table to link the \ac{pe} file with libraries, or data (e.g., strings 
used by the code section).
Finally, additional special sections, such as a debug section, might be 
added at the end of a \ac{pe} file and are often referred to as the 
overlay.

\subsection{Windows Malware Detection}

The research on new methods for malware detection based on machine
learning techniques is still a lively field in security, as these
approaches have the ability to derive malicious patterns and relations
from large datasets
automatically~\citep[e.g.][]{KolMal06,PerLanLee08,ArpSprHubGas+14,MarOnwAndCri+17}.
Due to these capabilities, learning-based methods can better cope with
the increasing number of emerging malware than traditional,
signature-based approaches.

As a result, researchers have proposed a large number of different
learning-based methods for malware detection throughout the past few
decades. These methods can be broadly divided into approaches based on
\emph{static}~\citep{KolMal06,PerLanLee08,JanBruVen11} and
\emph{dynamic}~\citep{BaiObeAndMao07,RieHolWilDue+08,BayComHlaKrue09,KolComKruKir09,LanBalKruChr10}
analysis. For instance, an early approach that applies machine
learning to detect malicious \ac{pe} files has been presented by
\citet{KolMal06} in 2006. Their method uses boosted decision trees
trained on statically extracted byte n-grams of a \ac{pe} file.
Similarly, \emph{BitShred}~\citep{JanBruVen11} clusters n-gram
features to perform large-scale malware triage.
In contrast to these approaches, which are mainly based on static
analysis, there are also a number of approaches that rely on
dynamic analysis~\citep{BayMosKru+06, WilHolFre07, BaiObeAndMao07, 
BayComHlaKrue09,KolComKruKir09,LanBalKruChr10}.
Due to the prediction time constraint of five seconds, we refrained from
incorporating them into our defense.

A recent learning-based detection method for Windows malware is the
so-called \ember classifier, which uses handcrafted features and
yields high detection rates, even outperforming related, deep
learning-based approaches~\citep{AndRot18}. The \ember classifier
builds the foundation for some of the defenses presented in this paper 
(see \autoref{sec:defenses}) due to the provided data.
In the following, we describe some of the details on the \ember 
classifier and the features it uses.

\paragraph{The \ember classifier} The \ember classifier has been
trained on a large corpus of Windows PE files, containing more than 1
million applications in total. Unfortunately, the binaries are not
publicly available due to legal restrictions. However, the authors of
\ember have released all features extracted from this corpus, enabling
other scientists to reproduce the results and use the data for further
research. 

The \ember feature set comprises eight groups of raw features, which
can be broadly divided into two different feature categories:
\emph{parsed features} and \emph{format-agnostic features}. The parsed
feature sets include various features extracted after parsing the
file, including basic information derived from the
headers as well as imported and
exported functions. In contrast, the format-agnostic features can be
extracted without parsing the file. These feature sets include simple
statistics about printable string characters and byte histograms.

To derive distinct patterns for malicious and benign applications, the
extracted feature sets are first mapped into a common feature space.
Afterward, a \ac{gbdt} model is trained on the feature vectors
with \lgbm~\citep{Frie00}.

\subsection{Evasion Attacks With Structured Data}\label{subsec:evasion}
Our focus in the competition lies on evasion attacks against learning 
methods, where an adversary tries to manipulate a sample,
such that it is misclassified. %
Unlike the commonly studied image domain, programs are a structured
input and any manipulation needs to keep the functionality. The action
space is limited, \ie bytes in a PE file cannot be changed arbitrarily
in general. To create real adversarial examples for malware, an
adversary needs to consider multiple constraints, such as defining
possible manipulations, keeping the functionality, and preserving the
inconspicuousness~\citep[see][]{PiePenCor+20}. 
In general, the attacker faces the challenge that problem and 
feature space have no one-to-one 
correspondence~\citep[see][]{QuiMaiRie19}. The PE file 
needs to be changed in its input or problem space, but the machine 
learning method operates in a feature space. Consequently, finding the 
optimal solution while keeping all constraints is challenging. 
It has a direct impact on the attacker's capabilities.

Although various rather advanced attack strategies are already 
discussed~\citep[e.g.][]{QuiMaiRie19, PiePenCor+20, XuQiEva16}, we 
identify that current attacks often use rather simple, yet considerably 
effective weak spots in learning-based systems. These weak spots 
simplify a real-world attack and need to be considered~as~well.

Finally, we note that various evasion techniques against malware 
detection exist that do not directly target machine 
learning~\citep[e.g.][]{WagSot02,MosKruKir07,SonLocStaKerSto07}. For
instance, adversaries can obfuscate the control flow, data location or 
data usage and so deceive the feature extraction from a 
static analysis~\citep{MosKruKir07}.

\section{Attacker's capabilities}\label{sec:attacks}
Our learning-based system needs to detect Windows \ac{pe} malware
while an attacker is targeting the learning-based system itself. It is 
thus important to analyze the attack surface first. In the following, 
we examine the attacker's capabilities being relevant for the 
competition before introducing our learning-based system as a defense 
in the next section.

\begin{figure}[]
	\centering
		\includegraphics{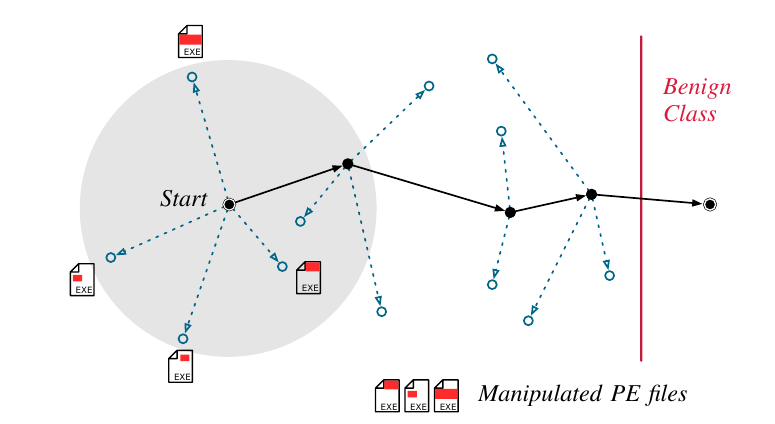}
	\caption{An evasion attack against a learning-based system
		consists of two components: different modifications and an 
		algorithm to find a suitable sequence of them towards the 
		benign class.}
	\label{fig:attackstrategy}
\end{figure}

An adversary has to change the features to create adversarial examples. 
As described in \autoref{subsec:evasion}, the challenge is to
manipulate a \ac{pe} file without compromising its functionality.
Such an attack can be divided into two components, as
\autoref{fig:attackstrategy} illustrates: (i) the file modification
(where and how a \ac{pe} file can be changed), and (ii) the attack
algorithm (how to combine modifications for evasion). 
Both need to be combined accordingly, such that the adversary finds a 
sequence of modifications that lead her to the benign class.

\subsection{File Modification}
A prevalent modification strategy in the literature and previous
competitions against learning methods is to add unused content in
areas that are not relevant for the functionality of a
program~\citep[e.g.][]{DemCouBig+20,web:Fle19}.  In particular,
adversaries exploit the \emph{semantic gap}, that is, the discrepancy
between extracted features and the actual processed part of a PE
file~\citep{JanShm12}.
Although this is a rather brittle attack procedure, it has a
considerable advantage: Changing or adding content in unused areas
relieves the adversary from implementing modifications that keep the
program functionality.  At the same time, the classifier is influenced
by these areas. Therefore, the semantic gap as a weak spot of a
learning-based system can have a considerable impact on its security.

\begin{figure}[]
	\begin{subfigure}[b]{.3\linewidth}
		\centering
	\includegraphics{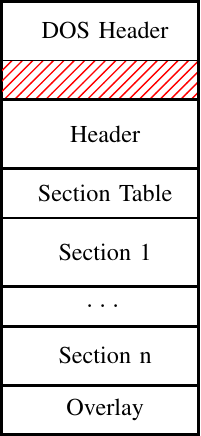}
		\caption{Extended header}
		\label{fig:pe_semantic_gapA}
	\end{subfigure}
	\begin{subfigure}[b]{.3\linewidth}
		\centering
	\includegraphics{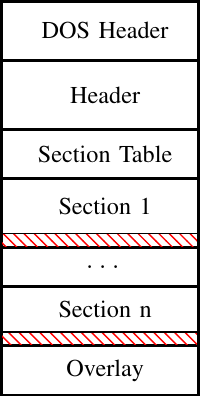}
		\caption{Filled slack space}
		\label{fig:pe_semantic_gapB}
	\end{subfigure}
	\begin{subfigure}[b]{.3\linewidth}
		\centering
	\includegraphics{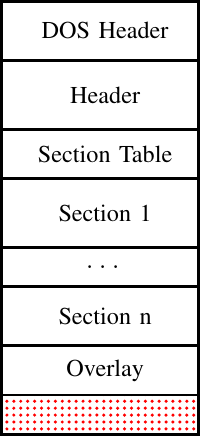}
		\caption{Extended overlay}
		\label{fig:pe_semantic_gapC}
	\end{subfigure}
	\caption{Semantic Gaps: Different unused locations that do not 
		influence the program functionality, but the learning-based 
		system. }
	\label{fig:pe_semantic_gap}
\end{figure}

In the context of PE files, an adversary can exploit or create unused 
areas at multiple locations~\citep{DemCouBig+20}. As 
Figure~\ref{fig:pe_semantic_gap} highlights, it is possible to 
enlarge the DOS header, to fill the slack space at the end of each 
section, or to append bytes to the overlay, \ie to the end of the 
\ac{pe} file. Another possibility is to add a new unused section that 
the adversary can fill~\citep{AndKhaFil+18}. 

A rather simple, yet effective way is then to inject content from
\emph{benign} samples into the unused areas to overload the malicious
traits~\citep{web:Fle19}.  For instance, adding a large number of
strings extracted from a benign Windows file can successfully evade
the default Ember model~\citep{web:Fle19}. More general, this attack
represents a mimicry attack that manipulates an input such that it
mimics the characteristics of a particular target 
class~\citep{FogShaPerKolLee06}.

Another way is to rewrite the features. An attacker can be expected to
try exploiting weak spots in the features first. For example, the
model may focus on a few features only, or the model uses features
that are simple to change, such as the timestamp in a \ac{pe} header.
\citet{AndKhaFil+18} introduce further modifications, such as
packing/unpacking, rewriting section names, and manipulating the
header checksum.

\subsection{Attack Algorithm}\label{subsec:attalg}
An attacker can proceed in different ways to find the combination of 
modifications that evade the learning-based system. Recall that we 
operate in a black-box scenario in the competition. 
Still, the black-box access is enough to find promising 
modifications by \emph{iteratively} querying the 
system~\citep[e.g.][]{DanHuaCha17, XuQiEva16, QuiArpRie18, 
	QuiMaiRie19, AndKhaFil+18}. For instance, attacks based on 
	Monte-Carlo tree search can evaluate the impact of modifications 
	multiple steps ahead by using a search 
	tree~\citep[see][]{QuiMaiRie19}.

It is reasonable to expect that an attacker can also learn an own 
local surrogate model from a similar training dataset or feature 
set due to some domain knowledge~\citep{BigRol18}. This can reduce the 
number of queries considerably.
Identified weak spots in a surrogate model may allow an attacker to 
create evasive samples that transfer to the original model under 
attack~\citep{RndLas14, PapMcDGoo16b}. In our case, multiple of our 
models are based on the \ember feature set. We can therefore assume 
that an adversary also tries promising directions on an own \ember 
model and verifies them with the black-box access to our defense. This 
was indeed the case for attackers in the 
competition~(see~\autoref{sec:eval}).

With a surrogate model, an attacker can use white-box attacks to 
compute evasive samples. Computing a gradient, for instance, has the 
advantage of finding the direction towards the benign class. Still, the 
gradient from the feature space needs to be mapped back to a valid file 
modification in the problem space.
\citet{KolDemBig+18} examine such a method to create a valid byte
modification. Yet, this attack builds on the semantic gap, so 
that the byte modification does not need to consider the functionality 
of the \ac{pe} file.

\section{Defenses}\label{sec:defenses}

\begin{figure}
	\centering
	\includegraphics{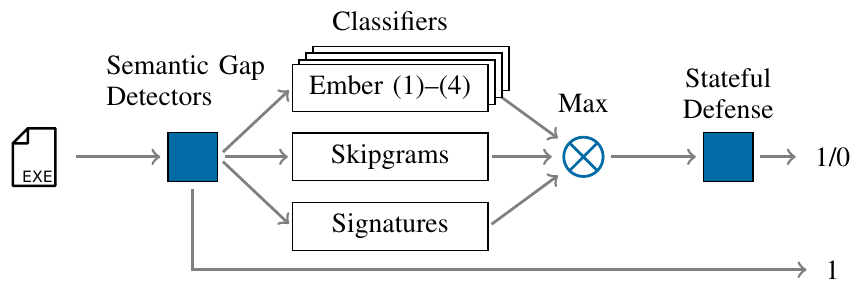}
	\vspace{-0.3cm}
	\caption{Overview of our learning-based system that detects Windows 
		\ac{pe} malware and considers an attacker targeting the 
		learning-based system itself.}
	\label{fig:overview_defenses_small}
\end{figure}

Equipped with a basic understanding about the attacker's capabilities,
we are ready to proceed with our developed approach \defnameit. 
The overall design concept is diversification. Our system consists of 
multiple defenses, each addressing different attack strategies
outlined before. A \ac{pe} file is classified as malware if \emph{any} 
of the system's components considers it as malicious. Therefore, an 
attacker needs to exploit weaknesses in all components in order to 
successfully trick the system. \autoref{fig:overview_defenses_small} 
gives an overview of the defenses and their combination.
In the following, we describe the concept of each defense in more 
detail. \autoref{table:defensesoverview} provides a summary of our 
approach.

\subsection{Malware Classification}
We start with introducing three different approaches to detect malware.
This ensemble increases the diversity, so that not only the malware 
detection can be improved, but also an attacker has to evade different 
feature sets at the same time.

\paragraph{Ember-based Model}
Our first and main defense is based on \ember~\citep{AndRot18} due to 
its high detection performance and the provided features from a large 
corpus of data. 
Yet, we exclude the header feature group in all models to reduce the 
attack surface. 
Preliminary tests show that these features, such as a timestamp, can
be easily manipulated by an adversary. This, in turn, can considerably
affect the classification performance.  Moreover, we rely on various
regularization strategies of the learning algorithm to prevent the
model from focusing on a few features only.

We train a \ac{gbdt} model using xgboost~\citep{CheGue16}.
In particular, our default model is trained on all \ember
features extracted from the 2017 corpus, except for the header
features. Also, we add multiple variations of this model to our
ensemble. In this way, we can mitigate various attack strategies and 
still use the \ember feature set from the large corpus of data.

Our first variation is to truncate the input by using the virtual
size, that is, the binary size after mapping the file into memory.
Only the bytes up to the virtual size of a \ac{pe} file are passed to
the feature extraction. The intention is to reduce the impact of
attacks that add malicious bytes to the overlay. Furthermore, we
remove strings in the byte stream passed to the byte histogram and
byte-entropy histogram feature set from \ember. The idea is to reduce
the impact of adding strings from benign samples on the histogram
distribution.

Our second variation uses only a reduced feature set from \ember. In
particular, we only consider the section, import, export, general
file, and string groups. In the latter group, we do not use the
printable string histogram. In this way, we remove all histogram parts
that are vulnerable to feature addition attacks. Moreover, the data
directory group is not considered as well, as preliminary tests showed
its vulnerability against feature manipulation.  

Our third variation uses the same features as the second variation,
but is trained on the 2018 corpus. It adds new knowledge about
more recent malware. In principle, we could train a \ac{gbdt} model on 
the combined 2017 and 2018 dataset, but we learned two models to reduce
training time.

\begin{table}
	\centering
	\footnotesize
	\renewcommand{\arraystretch}{1.25}
	\caption{
		Overview of implemented defenses
	}
	\label{table:defensesoverview}
	\begin{tabularx}{0.49\textwidth}{L{1.9cm}L{1.7cm}X}
		\toprule
		\head{Defense} & \head{Type} & \head{Description} \\
		\midrule
		Semantic gap detector & Slack space & Scan for filled slack 
		spaces 
		\\
		& Overlay & Check if overlay is too large compared to the file 
		size 
		\\
		& Duplicates & Check if same sections are used\\
		\midrule
		Classification & \ember-based GBDT~models
		& 
		\textit{Default}: Without PE header 
		features, trained on 2017 corpus \\
		&  & \textit{1st variation}: Without PE header, string 
		stream removed, input truncated, 2017 corpus \\
		&  & \textit{2nd variation}: Reduced feature set, input 
		truncated, 2017 corpus \\ 
		&  & \textit{3rd variation}: Reduced feature set, input 
		truncated, 2018 corpus \\ 
		\cmidrule{2-3}
		& Skipgram model & Monotonic GBDT model based on skipgrams 
		(3-skip 3-grams)\\
		\cmidrule{2-3}
		& Signature-based~model & Detection based on Yara rules for 
		popular malware families\\
		\midrule
		Stateful defense & & Monitor incoming queries for patterns 
		indicative of evasion\\
		\bottomrule
	\end{tabularx}
\end{table}

\paragraph{Monotonic Skipgram Model}
Although the previous variations mitigate the impact of feature 
addition attacks, our next model is invariant to such attacks by design.
To this end, we use the concept of monotonic 
classification~\citep{IncTheAfr+18}. With a monotonic model, an 
increase in feature value can only increase the malware score, so that 
more benign features cannot lower the classification score.
Unfortunately, this leads to a lower detection performance in 
general~\citep{IncTheAfr+18}. Still, monotonic models can serve as 
an additional line of defense against feature addition.

\begin{figure}[t]
	\centering
	\includegraphics{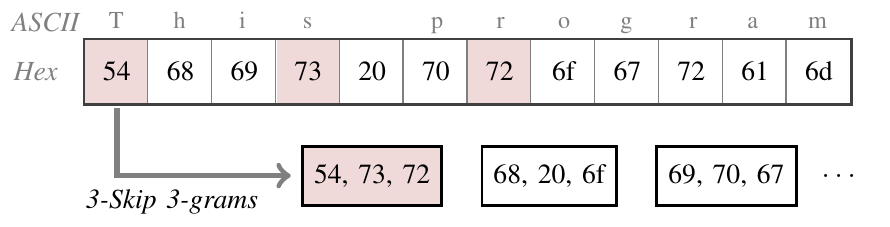}
	\vspace{-0.3cm}
	\caption{Illustration of skipgrams extracted from a byte sequence. 
	The n-gram length and gap size are 3, respectively.}
	\label{fig:skipgrams}
\end{figure}

As features, we use skipgrams extracted over the bytes of a \ac{pe} 
file. In principle, skipgrams are n-grams, but with gaps between each 
token. \autoref{fig:skipgrams} underlines the concept.
The absolute count of each skipgram is used, so that the model is 
based on the presence of features~\citep{CesBotGom+19}. 
The relative frequency, for example, is not used. A normalization
allows decreasing monotonic malware features by adding 
other features.
We train a monotonic \ac{gbdt} model with xgboost based on an own 
collected dataset of \num{22000}~benign and malicious \ac{pe} files.

\paragraph{Signature-based Model} The third component is based on 
Yara rules\footnote{\url{https://github.com/Yara-Rules/rules}} that 
capture the characteristics of well-known malware as detection 
signatures. In particular, we extract all matched rules on our 
collected \ac{pe} file dataset and keep only those rules that are 
solely present in malware. The system assumes malware if any of the 
kept rules is matched. Note that these rules have the disadvantage of 
being manually crafted signatures. Although they provide only a 
small increase in detection performance, they serve as a backup by
capturing malware patterns directly.

\subsection{Semantic Gap Detectors}

As an additional line of defense, we try to reduce the semantic gap,
so that the adversary is forced to perform more complicated changes
than adding features in unused areas. To this end, we
implement detectors to check if an attacker exploits a semantic gap.
We consider the following three approaches. 
Our \emph{slack space} scanner assumes an attack if the space between 
sections is filled with non-zero bytes. The reason is that the slack 
space of benign samples is expected to just be filled with zeros, such 
that the sections match with the file alignment. 
Our \emph{overlay} scanner computes the ratio of the overlay to 
the overall file size. A considerably high ratio indicates that an 
attacker may have appended bytes to the overlay. 
Finally, our \emph{duplicate} scanner detects an attack if two or more
duplicate sections are present, excluding empty sections. We assume 
that benign files have rather unique sections.
This idea reduces the attack surface by preventing the 
attacker from adding the same benign content repeatedly. 

Note that these are rather simple heuristics and only cover %
prevalent semantic gaps. Still, they raise the bar for a successful
attack, which has also been confirmed throughout the competition
(see~\autoref{sec:eval}). Our detectors motivate a more systematic
analysis to close the semantic gap between extracted features and the
processed parts of a \ac{pe} file, enforcing the attacker to change
used code then.

\subsection{Stateful Defense}
The attackers in the competition operate in a black-box scenario with
access to a binary classification output. Thus, they may need to use 
iterative attack algorithms as described in \autoref{subsec:attalg}. 
The attackers send specifically crafted samples and observe the 
respective binary output over multiple iterations.
This, in turn, motivates a stateful defense that monitors incoming
sequences of queries for evasion patterns. 

For the sake of simplicity, we adapt the stateful nearest-neighbor
detector already given by the competition. The overall idea is to
check if a similar, previously submitted file has been considered
malicious. In particular, the detector saves each sample classified as
malware in a history buffer. 
For a sample under investigation, the detector checks if it is 
sufficiently similar to a previous malicious one. If the distance to 
the nearest neighbor from the history buffer is smaller than a 
defined threshold, the detector assumes an attack.
To obtain the distance between two \ac{pe} files, we calculate the 
$L_1$ distance on the histogram and byte-entropy features from \ember.
In contrast to the competition's solution, we do not truncate the 
\ac{pe} file to limit the number of sections or bytes per section.

Another approach is to save any input (\ie benign or malicious) in the
history buffer to check if the system is probed with a sequence of
similar inputs. We decide against this strategy to lower the risk of
our solution being rejected due to a high false positive rate, as this 
rate must not exceed~\perc{1} throughout the
competition. The competition organizers may regularly submit benign
samples to validate the solution, so that benign samples once stored
in the history buffer can create false positives in these checks.

We finally note that stateful defenses can considerably impede 
iterative attacks, and thus should be explored further. Different 
stateful strategies have already been successfully applied in 
multimedia security that are transferable to machine
learning~\citep[see][]{QuiArpRie18}.

\subsection{Combination of Defenses}
Our system consists of a diverse set of defenses, including different
machine learning models, semantic gap detectors, and a stateful
defense. Figure~\ref{fig:overview_defenses_small} shows this 
combination. The input file is first passed to the
semantic gap detectors. If they detect a malicious change, the system
classifies the input as malware. Otherwise, we use our ensemble of
classifiers, including the four Ember-based \ac{gbdt} models, a
monotonic skipgram model, and a signature-based classifier. We use max 
voting to combine all predictions,
so that an attacker has to evade all models for a successful attack.
With a positive malware classification, we pass the input to the
stateful defense in order to save it in the history buffer, and return
the malware classification. With a benign prediction from the
ensemble, the stateful defense finally checks if the input sample is
too similar to a previous one. If so, the system predicts malware.

\section{Competition Results}\label{sec:eval}

In the competition, five attackers participated. They attacked three 
defense solutions: (i) our defense, (ii) the one from the team 
'needforspeed'~\citep{web:CesBot20}, and (iii) an \ember-based 
baseline given by the competition~\citep{web:GithubContest}.
To be accepted to the attacker challenge, a defense has to detect 
regular malware reliably on an unknown holdout set (see 
\autoref{subsec:competition}).  

Each submitted attack sample is checked if it evades any defense before 
its functionality is validated, \ie it has the same behavior as the 
original malware sample.
A defense is ranked according to its relative evasion 
rate, that is, the fraction of all validated samples that evade the 
particular defense.

\paragraph{Results}
\autoref{table:defenderchallenge} shows the results from the defender 
challenge. Our solution got the first place with a relative evasion 
rate of \perc{23}, compared to \perc{58} and \perc{94} for the other 
defenses. Thus, our defense considerably increases the 
robustness against attacks, and withstands different attackers.

\begin{table}
	\centering
	\footnotesize
	\renewcommand{\arraystretch}{1.25}
	\caption{
		Results from defender challenge in terms of evasion rate
	}
	\label{table:defenderchallenge}
	\begin{tabularx}{0.45\textwidth}{L{0.4cm}XX}
		\toprule
		\head{Rank} & \head{Defense} & \head{Relative Evasion Rate} \\
		\midrule
		1. & \textbf{Our defense \defname} & \textbf{\perc{23}} \\
		2. & needforspeed~\citep{web:CesBot20}  & \perc{58} \\
		3. & \ember-based baseline~\citep{web:GithubContest}  & 
		\perc{94} \\
		\bottomrule
	\end{tabularx}
\end{table}

\paragraph{Analysis}
A closer look on the attacks provides further insights.  Only the team
`needforspeed'~\citep{web:CesBot20} succeeded against our system for 
all 50 samples. This team uses a dropper with a single-byte XOR 
encryption and Base64 encoding to hide the payload. Our system cannot 
cope with these kinds of obfuscations, as it solely builds upon 
features extracted through static analysis, making it difficult to 
reconstruct the encrypted content.

Further attackers are less or not successful. 
Using self-signed certificates~\citep{web:Wun20} succeeds only for 
\perc{18} of the files. Adding content from benign files to the overlay 
or a new section guided by genetic programming~\citep{DemBigLag+20} 
leads to no successful sample against our defense.

All in all, most attackers seem to have constructed their samples with 
no concept of learning in the back. For instance, they target the 
static analysis by hiding the payload~\citep{web:CesBot20} or add
self-signed certificates~\citep{web:Wun20}. Thus, current attacks are
\emph{so far} rather expert-driven attacks than systematic attacks on a 
learning model.
Finally, the competition also underlines that an attacker can 
exploit some knowledge even in a black-box scenario. The adversaries
used own local models and domain knowledge first to enhance and test 
their attacks before making queries~\citep[e.g.][]{web:CesBot20}. 
This explains the rather small number of \num{741} API queries of the 
leading attacker~`needforspeed'. 
Yet, the more robust a learning-based system becomes, the more we can 
expect adversaries to systematically send queries in the 
future.
\\

\section{Conclusion}
\label{sec:conclusion}

Modern learning-based systems for malware detection  do not only have
to identify malware reliably. Also, they should be robust against
attackers that target the system itself.  The Microsoft Evasion
Competition is an excellent opportunity for security researchers to
analyze and reduce the attack surface of state-of-the-art
learning-based detection methods.

Our defense for the competition is based on diversification to consider 
multiple attack scenarios.
In particular, we address the semantic gap, use various classification 
models, and apply a stateful defense for iterative attacks. 
Our defense shows that existing machine learning methods can be 
hardened against attacks by analyzing the attack surface thoroughly and 
implementing concepts from adversarial learning.

However, we also find that our system is still prone towards
obfuscation methods, as it solely builds upon features
extracted throughout a static analysis. While many attack strategies
have been successfully fended off by our model, including the
exploitation of the semantic gap, circumventing the system by
encrypting the malicious payload has been successful. To counter such
attacks, extending the system with a dynamic analysis should be
considered in the future.

\section*{Acknowledgment}
We acknowledge funding by the Deutsche 
Forschungsgemeinschaft (DFG, German Research Foundation) under
the research grant \mbox{RI 2469/3-1}, and by the German 
Ministry for Education and Research as BIFOLD - Berlin Institute for 
the Foundations of Learning and Data (ref. \mbox{01IS18025A} and ref 
\mbox{01IS18037A}).
Furthermore, we would like to thank Christian 
Wressnegger for supporting us with additional data.

{\footnotesize
	\interlinepenalty=10000

}

\end{document}